\documentstyle[12pt,epsf]{article}

\def\be{\begin{equation}}
\def\ee{\end{equation}}
\def\bea{\begin{eqnarray}}
\def\eea{\end{eqnarray}}
\def\ba*{\begin{eqnarray*}}
\def\ea*{\end{eqnarray*}}
\begin{document}
\newcommand{\sheptitle}
{ What Can WMAP Tell Us About The Very Early Universe?
New Physics as an Explanation of Suppressed Large Scale Power
and Running Spectral Index}

\newcommand{\shepauthor} {Mar Bastero-Gil$^{ (1)}$, Katherine Freese$^{
(2)}$ and Laura Mersini-Houghton$^{(3)}$} \newcommand{\shepaddress} {$^1$
Centre for Theoretical Physics, University of Sussex \\ Falmer,
Brighton BN1 9QJ, United Kingdom \\ $^2$ Michigan Center for
Theoretical Physics, Ann Arbor, MI 49109\\$^3$
Department of Physics, Syracuse University }

\newcommand{\shepabstract} 
{The Wilkinson Microwave Anisotropy Probe
  microwave background data may be giving us clues about new physics
  at the transition from a ``stringy'' epoch of the universe to the
  standard Friedmann Robertson Walker description.  Deviations on
  large angular scales of the data, as compared to theoretical
  expectations, as well as running of the spectral index of density
  perturbations, can be explained by new physics whose scale is set by
  the height of an inflationary potential.  As examples of possible
  signatures for this new physics, we study the cosmic microwave
  background spectrum for two string inspired models: 1) modifications
  to the Friedmann equations and 2) velocity dependent potentials.
  The suppression of low ``l'' modes in the microwave background data
  arises due to the new physics.  In addition, the spectral index is
  red ($n<1$) on small scales and blue ($n>1$) on large scales, in
  agreement with data.}

\begin{titlepage}
\begin{flushright}
hep-ph/\\
SU4252-780/\\
SU-GP-03/5-1 /\\
MCTP-03-20\\
%\today
\end{flushright}
\vspace{.1in}
\begin{center}
{\large{\bf \sheptitle}}
\bigskip \medskip \\ \shepauthor \\ \mbox{} \\ {\it \shepaddress} \\
\vspace{.5in}
%{\bf Abstract}

\bigskip \end{center} \setcounter{page}{0}
\shepabstract

%\baselineskip 24pt

%\maketitle
%\begin{abstract}
%\end{abstract}
%{\Large \bf Equation of state }

\begin{flushleft}
\hspace*{0.9cm} \begin{tabular}{l} \\ \hline {\small Emails:
mbg20@pact.cpes.susx.ac.uk, ktfreese@umich.edu,

l.mersini@phys.syr.edu} \\

\end{tabular}
\end{flushleft}

\end{titlepage}
%%%%%%%%%%%%%%%%%%%%%%%%%%%%%%%%%%%%%%%%%%%%%%%%%%%%%%%%%%%%%%%%%%%%%%%%%%%%%%%%%%%%%
\section{ Introduction}

The high precision of observational cosmology has placed tight
constraints on various cosmological parameters and so far seems to be
in excellent agreement with the\footnote{Concordance model is a
spatially flat Universe with adiabatic, nearly scale invariant initial
fluctuations. The most popular variant is a $\Lambda$CDM model.}
``concordance model''. The Wilkinson Microwave Anisotropy Probe (WMAP)
is an important experiment in this field oriented at measuring the
anisotropies in the Cosmic Background Radiation (CBR). Its recent
findings~\cite{wmap1,wmap2, wmap3}, especially the suppression of the
spectrum at large angular scales and the running of the spectral
index, have generated a source of great excitement and
speculation~\cite{berera,tegmark,esftathiou,uzan,alinde,linde,kofman,cline,cst}.           
Unfortunately it is difficult to isolate the findings of suppression  
of low-$l$ modes from cosmic variance limitations\footnote{ 
See for example Ref. \cite{berera2} for an attempt
to understand the cosmic variance limitations in terms of its
implications on the k-space  power spectrum.}.  However, we here
investigate the possibility that the lack of signal in the temperature
angular correlation function on angular scales $\theta \ge 60^0$ is a
hint of new physics.  In addition, there is some indication in the
data that the spectral index runs from red on small scales to blue on
large scales as compared to a scale invariant spectrum, although the
statistical significance of these results is not entirely clear
\cite{seljak,wang,barger}.  Both of these effects in the data, the
suppression of large scale power and the running of the index, can be
explained by new physics.

Some of the recent speculations which attempt to explain the lack of
power around $\theta \geq 60^0$ involve a finite-size universe with
non trivial topology~\cite{starkman,levin,tegmark} or a closed
universe~\cite{esftathiou,uzan,alinde}. It is too early to conclude whether
these ideas would accommodate the rest of the cosmological data that
fit so well with the ``concordance model''~\cite{glenn}. 
Another proposed explanation for the suppressed power on large scales
involves double inflation with a period of chaotic inflation
followed by new inflation due to a single potential \cite{yoko,feng}.

We take the Universe to remain flat and topologically
trivial, and take the point of view that the suppression of the
correlations at low multipoles $l\simeq \frac{180}{\theta}$ is
providing us with clues about the initial conditions of inflation.
More precisely our claim is that from the low ``l'' suppression of the
signal, we could be learning about the {\it initial conditions} for
the part of inflation that gives rise to observables in structure
formation and in the microwave background. The initial conditions for
the observable universe could {\it be determined by the new physics},
which is the relevant theory valid beyond, and around the cutoff scale
$E_c$ of effective low energy theories.

Inflationary cosmology \cite{guth} was proposed as a solution to the
horizon, flatness, oldness, and monopole problems of the standard
cosmology.  An early epoch of superluminal expansion explains why the
universe looks so homogeneous and flat today. We here consider
potential-driven models of inflation, in which a potential dominates
the energy density of the universe and causes the universe to
accelerate.  In particular, in models where the height of the
potential $V$ is determined by the GUT scale, roughly 60 efoldings of
inflation are required to produce a universe that is homogeneous out
to the horizon scale.  It is only these last $60$ e-foldings of
inflation that give rise to the observed structure at present, and the
scale of our present Hubble horizon $H_0$ would correspond roughly to
the $60^{th}$ e-folding before the end of inflation.  We here consider
models in which physics beyond the inflationary scale (e.g. GUT scale)
is modified by the introduction of new physics; this new physics can
leave imprints in the CMB at scales near and above the horizon.

Above the $60^{th}$ e-folding the theory is described by new physics.
This is the regime corresponding to all the energy scales above $E_c$,
the cutoff scale of the effective low energy
theories. According to this point of view it then follows that the low
``l'' feature in CBR may originate from the new physics taking place
around the initial time of inflation. 

The very low $l$ modes in the CBR spectrum at present time,
correspond to very large wavelength modes of the order of the current
Hubble scale, $H_0 ^{-1}$.  Around the e-folding time corresponding to
the cutoff scale, these were the first modes to cross the horizon and
are the last ones to re-enter at present.  Since these modes have been
superhorizon sized between inflation and now, they have not been
contaminated by the later evolution of the Universe.  This means that
the information we get from these modes in the primordial spectrum can
contain and probe {\it only} very early-time effects, which we refer
to as {\it pristine} information. For this reason we could attribute
the new feature observed in the spectrum at low $l$ to the initial
conditions of inflation.

%%%%%%%%%%%%%%%%%%%%%%%%%%%%%%%%%%%%%%%%%%%%%%

Since we use quantum field theory (QFT) and general relativity (GR) as
our conventional tools to study and describe inflation, it does not
make sense to study inflation and the range of e-foldings
corresponding to scales above the cutoff with these tools, as they
cease to be valid. Due to the new physics occurring at scales above
$E_C$, we can not use GR and QFT equations to draw any credible
conclusions. For this reason, the e-folding corresponding to the
cutoff scale {\it determines} the ``onset'' time for the conventional
inflation.
 
New physics is expected to ``kick in'' and be important above $E_c$,
for example string theory. Let us assume that slightly above and
around $E_c$ there was a period where some kind of unknown transition
took place. This transition is needed to smoothly bridge the end of
the string regime at (which we are taking here to be the fundamental
high energy theory), to the epoch of the conventional ``lower'' energy
inflation $V < E_c$, as described by QFT and GR formalisms.

As specific examples of new physics above the scale $E_c$, we consider
two modifications to the formalism of standard cosmology: 1) the
effective Friedmann equation receives stringy
corrections~\cite{binetruy,flanagan,csaki,laura}; or 2) the inflaton
potential is velocity dependent due to relative brane
motion~\cite{shiu-tye}. Certainly, once inflation starts, after a few
e-folds our patch of spacetime has inflated sufficiently, that the
``string era'' becomes an irrelevant description and we enter/recover
the conventional cosmology period of the standard model (see, e.g.,
\cite{afk}).  That is to
say, that GR and QFT, soon after the transition from the string era,
should be recovered in the low energy limit.  

As our first example, we consider modifications to the Friedmann equation,
\be
\label{eq:mod}
H^2 = {8 \pi \over 3 M_{P}^2} \left(\rho + f[\rho^\alpha]\right), 
\ee
where $\rho$ is the energy density, $\alpha$ is a parameter, 
and $f$ indicates the modification term to the Friedmann equation. The
second term is important in the regime $\rho \geq \sigma$
where $\sigma$ is an energy scale, introduced by the underlying string model, to be defined below.  
In particular, we focus on the model of Randall and Sundrum
(hereafter RSI) that was proposed as an explanation of the hierarchy
problem~\cite{rs}. Here our observable universe is a three-dimensional
surface situated in extra dimensions.  During the inflationary
epoch, the Friedmann equation on our 3-brane becomes
\be
\label{eq:mod2}
H^2 = {8 \pi \over 3 M_{P}^2} [\rho + \frac{\rho^2}{2\sigma} ], 
\ee
where $\sigma$ is the brane tension. Recall that in
the original Randall-Sundrum model the brane tension has to be
negative in order to ensure stability of the bulk\footnote{There have
  been many variations of the model since then. When bulk scalar
  fields are introduced~\cite{gw}, the stability condition on the
  brane tension is relaxed and $\sigma$ can be positive.}.  Previous
authors ~\cite{maartens,huey} have considered inflation in the
presence of such a second term, but focused on the case of a positive
tension $|\sigma| << V$, with $V$ the energy scale of the inflaton
potential.  They found that the extra contribution to the Hubble
expansion helps damp the rolling of the scalar field so that the
condition for slow-roll inflation can be met even for a steep
potential.  Here on the other hand, we are interested in a very
different regime, with a mass scale \be
\label{eq:choice}
|\sigma|^{1/4} \sim V^{1/4} \sim [ 10^{-3/4} M_{P} ]. 
\ee 
The reason for this choice for $\sigma$ is that, in many inflationary models,
the height of the potential is given by the GUT scale. Thus 
the second term in Eqn.~(\ref{eq:mod2}) will be important at the ``onset''
of inflation, $\phi_i \simeq O(M_{P})$, exactly for this choice of
the brane tension, $\sigma \simeq V$.  As shown by ~\cite{afg}, 
inflationary potentials must be flat in the sense that the ratio
of the height of the potential to the fourth power of the width
must satisfy
\be
\Delta V / (\Delta \phi)^4 \leq O(10^{-7}) ;
\ee
this statement quantifies what is often known as the ``fine-tuning''
problem in inflation.
For $\Delta\phi \sim M_{P}$, then one needs $V \leq 10^{-2} M_{P}^4$.

Such a parameter choice is plausible for the model of Randall and
Sundrum ~\cite{rs}. In RSI, the four and five
dimensional Planck scales $M_{P}$ and $M_5$ respectively are related via
\be
M_{P}^2 = (3/4\pi) \left({M_5^2 \over |{\sigma}|}\right) M_5 ,
\ee
where $\sigma$ is the brane tension and is negative.
With Eqn.~(\ref{eq:choice}) and taking $M_{P} = 10^{19}$ GeV,
we require 
\be
\label{eq:tune}
M_5^{3/2} \sim M_{P}/20 ,
\ee
a reasonable value.
Here we are studying a situation in which, due to ``stringy'' effects,
the effective Friedmann equation is modified above the  scale $\sigma$ for
the inflaton potential $V$ but returns to normal soon after.   

We will show that the density fluctuations well below today's horizon
scale are not significantly altered, while fluctuations on large
scales (near the horizon scale) are substantially suppressed.  The
perturbations produced on the large scales near the horizon are
produced close to the scale of new physics. The role of the new
physics is to cause the breakdown of the slow-roll approximation. At
the very earliest times, the kinetic energy of the inflaton field
dominated over the potential.  Hence at these early times, density
fluctuations were not produced in the usual way.  Eventually the
potential does come to dominate the energy density.  Once the
potential drops to the scale $V \sim \sigma$, slow roll inflation
ensues.  This transition happens roughly 60 efoldings before the end
of inflation.  Smaller scale fluctuations are produced during the
slow-roll phase of inflation, where the height of the potential is
fairly constant, so that the slope of the power spectrum is only
mildly affected.

This proposed explanation of the suppression of large scale power in
the WMAP data relies on the fact that the scale of new physics $E_c =
\sigma^{1/4}$ exactly corresponds to the height of the inflationary
potential at 60 efolds before the end of inflation. This is the only
fine-tuning that is required.  Once this energy scale is set, then it
is automatic that kinetic energy domination suppresses large scale
perturbations at exactly the right angular scale, namely just below
the horizon.  The fine-tuning need not be severe, e.g, in the sense
that the required ratio of four and five dimensional Planck scales in
Eqn.~(\ref{eq:tune}) is not unreasonable.

A similar proposal by \cite{linde} also invokes the failure of
slow-roll at 60 e-folds before the end of inflation.  In that paper,
the authors proposed a change in shape of a hybrid inflation potential
exactly at the right time to obtain the large scale suppression.  The
fine-tuning there is of a different nature.  The paper of \cite{linde}
requires the potential or the initial conditions to be carefully
chosen. In this paper, on the other hand, we do not require any
special features of the potential, which can indeed be quite ordinary.
Instead, the violation of slow-roll arises due to modifications e.g.
to the Friedmann eqn.  It would be interesting to consider the
question of how to differentiate observationally between these
different proposals.

In addition, the new physics changes the running of the spectral
index. We will show that an oddity of the data, namely that
the index is red ($n<1$) on small scales and blue $(n>1)$ on
large scales is automatically explained in the model we study.

We demonstrate in Section 2 how two types of modifications that
generically arise in string theory, those with modified Friedman
equation and those containing velocity dependent inflaton potentials,
may leave important imprints in the primordial spectrum.
We also show the result generated with the CMBFAST program,
that convert the primordial spectrum, $\delta_H$ where these
signatures are imprinted, to a present-day power spectrum, $P(k)$ and
accompanying anisotropies in the CBR. The resulting anisotropies for
two representative examples are then compared with WMAP data.  
These are summarised in Section 3.

\section{String correction effects on the primordial spectrum} 

Let us assume that, due to the underlying theory valid at high
energies, for example string theory, the Einstein equations receive
corrections. Some familiar examples can be found in brane-world
scenarios which generically either modify Einstein
equations ~\cite{binetruy,flanagan,csaki,laura,chung} or contain a velocity
dependent term in the inflaton potential~\cite{shiu-tye}. This velocity
dependent term, related to the relative brane motion, can also be recast as a
modification to the Friedmann equation and is calculable in string
theory\footnote{This term is also expected from post Newtonian
approximation in higher dimensional gravity.}. 
We know that at later times general relativity is a perfectly valid
theory and hence, we expect these corrections to be important and play
a role only at early times, during the transition from a string theory 
era to the onset of the standard brane-bound $3-dimensional$
inflation. Within very few 
e-foldings during inflation, our patch of spacetime becomes large
enough and these modifications become negligible. 

Here we are interested in addressing
the following issue: Do stringy corrections 
leave any imprint in the primordial power spectrum, and at what scales
would we expect to see this signature?

\subsection{The Spectrum for the Modified Friedmann Equation from Brane-Worlds}

Initially, let us consider the corrections to be of a general
form $f[\rho^{\alpha}]$, i.e., 
\be 
H^2= \frac{8 \pi}{3 M_P^2}
\left(  \rho  + f[\rho^\alpha]  \right )\,,  
\label{fried1}
\ee 
where $\rho$ is the inflaton energy density and $\alpha$ is a parameter.
Energy conservation requires
\be
\dot \rho + 3 H (\rho + p) =0  \,,
\ee
where $p$ is the pressure and
``dot'' denotes the derivative with respect to cosmic time. 
The acceleration equation for the scale factor $a$ is
\bea 
\frac{3 M_P^2}{4 \pi} \left( \frac{\ddot a}{a} \right)
&=& \frac{3 M_P^2}{4 \pi} \left( \dot H + H^2 \right) = -\left( \rho +
f[{\rho} ^{\alpha}] + 3p + 3 p_f \right) \,, 
\label{fried2}
\eea 
where $p_f$ would be an ``effective'' pressure contribution due to the
modification to the energy density in the Friedmann equation:
\be
p_f= (\rho+p) \frac{d f[\rho^\alpha]}{d\rho} - f[\rho^\alpha] \,.
\ee
Given the above setup, let us proceed in
calculating the primordial spectrum $\delta_H(k)$ for the case when
the Friedmann equation contains corrections of the form
Eqn.~(\ref{fried1}), that 
were introduced by the new high energy physics around the cutoff
scale. The symbol $\delta_H (k)$ denotes the primordial spectrum which gives
the amplitude of perturbations as the comoving wavenumber $k$ enters
the horizon. $P(k)$ denotes the present-day power spectrum giving us
the amplitude of perturbation at a fixed given time. They are related
by the transfer function $T(k)$ ~\cite{liddle}, for the present Hubble
value $H_0$, 
\be 
\frac{k^3}{2 \pi} P(k) = \left(\frac{k}{a H_0}\right)^4 T^2(k)
\delta_{H}^2(k)\,.  
\ee 

We will follow the standard approach to calculating the density
(curvature) perturbations created during inflation\footnote{We will neglect
Kaluza Klein modes.}, but will make the appropriate
modifications required by Eqn.~(\ref{fried1}).
During inflation, when the inflaton $\phi$ with energy $\rho\simeq V$
is the dominant one, the primordial power spectrum $\delta_H(k)$ is 
\bea
\delta_H &\simeq& \left(\frac{\delta \rho}{\rho + p }\right)_{k=aH} \,,
\eea
where, 
\bea
\delta \rho &\simeq& \delta \phi V^\prime \simeq \frac{H}{2 \pi}
V^\prime \,,\\
\rho + p &\simeq& - [ \frac{M_P^2}{4 \pi}\dot H +
f[\rho^\alpha] + p_f] \,, 
\eea 
with $V^\prime= dV/d\phi$. 
Henceforth, throughout the remainder of this section, all quantities in the
equations will be evaluated at horizon crossing, i.e.,
at $k=aH$.
Therefore, 
\be 
\label{eq:denom}
\delta_H \simeq \left( \frac{H}{2 \pi} \right) 
\left(\frac{- V^\prime}{\frac{M_P^2}{4 \pi}\dot H +
f[\rho^\alpha] + p_f}\right) 
\,.
\ee 
When the modification term $f[{\rho}^{\alpha}]$ 
is non-negligible, clearly $\delta_H$ ``feels'' the effect of ``the
stringy'' corrections. 

%%%%%%%% 
The spectral index $n_S$ can readily be found from the primordial
spectrum, Eqn.~(\ref{eq:denom}) 
\be
n_S -1 = \frac{d \ln \delta_H^2}{d \ln k}\,,
\ee
where the k-dependence of the spectrum is 
\be 
\delta^2 _H \propto \left( \frac{k}{k_S} \right)^{n_S-1}\,,
\label{modprimordial} \,.  
\ee
with $k_S$ a fiducial wavenumber depending on the dynamics of inflaton.

The latter expression for the primordial spectrum in terms of $k$
Eqn.~(\ref{modprimordial}), can be derived from Eqn.~(\ref{eq:denom})
by replacing in the solution for $\phi(t)$, the time dependence of the
field with the comoving momentum $k$, through the condition $k =a(t)
H$ and $a(t) = e^{N_e}$, with $N_e=$ number of e-folds. 
 
In the remainder of this section, we will present a rough estimate
of how density fluctuations may be modified due to the presence
of an energy cutoff in the spectrum given by $E_c$.  Then in the
following sections, we will present precise examples in which
we calculate exactly the resultant power spectrum. The rough
arguments in the remainder of this section give a good general idea
of our results.

Let us denote  the unmodified power spectrum by $\delta_H^0$ and
the unmodified spectral index by $n_S^0$, obtained by the limit
$f[\rho^{\alpha}] \rightarrow 0$. We can re-express the corrections in
the modified spectrum in Eqn.~(\ref{eq:denom}) and the running of the spectral
index as a function of the weighted wavenumber $k/k_S$.
%%%%%%%%%%%%%
\be 
\delta^2 _H = \delta^{(0) 2}_H 
\frac{1}{(1 + 4 \pi (f[\rho^\alpha]+p_f)/(M_P^2 \dot H))^2}  \propto
\left(\frac{k}{k_S}\right)^{n_S-1} 
\label{spectral} \,,  
\ee
with $\delta_H^{(0) 2}$ given in term of $n_S^{0}$
\be 
\delta^{(0) 2}_H \propto \left ( \frac{k}{k_S} \right)^{n^0_S-1}
\label{unmodprimordial} \,.  
\ee
Let us denote the term in brackets which is next to $\delta_H^{(0) 2}$
in Eqn.~(\ref{spectral}) by 
\be
g(\phi)= \frac{1 }{(1 + 
4 \pi (f[\rho^\alpha]+p_f)/(M_P^2 \dot H))^2} \,.
\label{gphi}
\ee 
Differentiating with respect to
$\ln k$ the correction to the spectral index is 
\be (n_S - 1) = (n^0_S
- 1) + \left( \frac{d \ln g(\phi)}{d \ln k} \right)= (n^0_S - 1) +
\delta n_S
\label{runspectral} \,.  
\ee
Therefore the correction term $\delta n_S$ is given by
\be 
\label{eq:correction}
\delta n_S = n_S-n_S^{(0)}=\left( \frac{d\ln g(\phi)}{d \ln k} \right) 
= \left( \frac{d \ln g(\phi)}{d \phi} \right) \left (\frac{d\phi}{d
\ln k} \right) = \frac{g(\phi)^\prime}{g(\phi)}\frac{\dot\phi}{H}
\,,  
\ee
where $d\phi/d\ln k = \dot\phi/H$ (evaluated at horizon crossing,
$k=aH$).
Note that since $g(\phi)$ is a function of $\phi(t)$ then the
correction $\delta n_S$ is generically a function of $k$. Its
k-dependence can be estimated in the same manner as we calculate
$\delta_{H}(k)$, by replacing the time dependence of the field in favour
of $k$ at $k = aH$. 

Formally, the $k$- dependence of $g(\phi)$ as a function
of $\delta n_S$, in a similar manner to $\delta_H$, is expressed by 
\be 
g(\phi) =
\left(\frac{k}{k_S}\right)^{\delta n_{S}(k)} \,. 
\label{gk}
\ee 
Depending on the model and the specific form of the modification
function $g(\phi)$, it should be noted that for certain values of $k$
the correction term, $\delta n_{S}(k)$ may become large and thus the
slow roll conditions may be violated, as illustrated in Sec.(2.3). 

But at $k \leq k_S$, we can recast Eqn.~(\ref{gk}) in the form
\be 
g(\phi) \propto \left [ 1 - e^{- (\frac{k}{k_S})^{\delta
n_S}} \right] . 
\ee 
The modified spectrum, with $\delta n_S$ given by
Eqn.~(\ref{eq:correction}), becomes  
\be 
\label{eq:exp}
\delta^2 _H = \delta^{(0) 2}_H
\left [ 1 - e^{- (\frac{k}{k_S}) ^{\delta n_S}} \right]
\label{spectral2} \,.  
\ee

\begin{figure} 
\epsfysize=9cm \epsfxsize=9cm \hfil \epsfbox{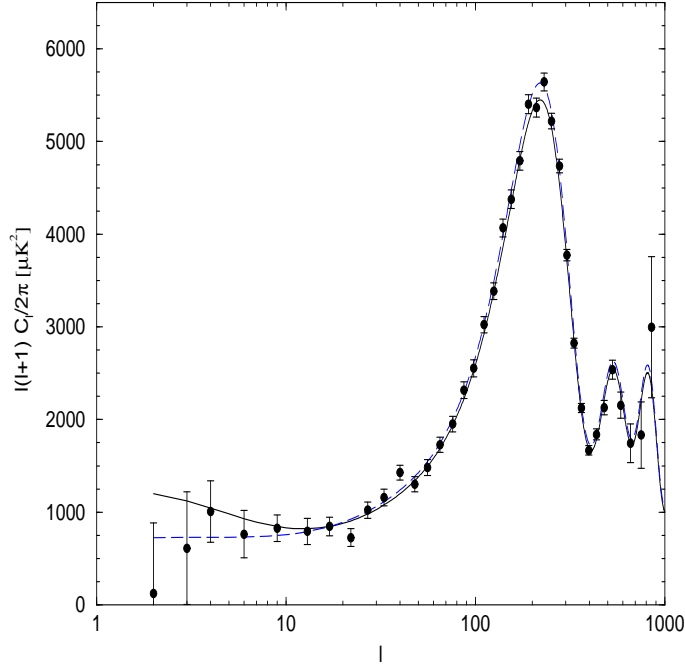} \hfil
\caption{\footnotesize CBR spectrum of anisotropies with the modified
primordial spectrum given in Eqn.~(\ref{eq:exp}), with
$\delta n_S=1$, and $k_S=0.0005\, Mpc^{-1}$. We also show for
comparison the curves for the standard 
spectrum Eqn.~(\ref{unmodprimordial}), with $n_S^{(0)}=0.99$. 
Cosmological parameters as given by the best fit model
of the WMAP collaboration.}
\label{plot1}
\end{figure}

In Fig. (\ref {plot1}) we show as an example the result generated
with the CMBFAST program, that convert the primordial spectrum to a
present-day power spectrum. We show for comparison the resulting
anisotropies for an unmodified primordial spectrum $\delta_H^{(0)}$,
with $n_S^{(0)}=0.99$, and the modification as given in
Eqn.~(\ref{eq:exp}), with the choice of parameters $\delta n_S=1$ and
$k_S=0.0005\, Mpc^{-1}$.  We allowed all the
parameters to be the conventional ones of a $\Lambda$CDM universe, 
with matter density $\Omega_M = 0.27$,  
vacuum energy density $\Omega_\Lambda=0.73$,  baryons $\Omega_b=0.046$, and
$H_0= 71 {\rm km s}^{-1} {\rm Mpc}^{-1}$. We note 
a very similar study to Fig. (\ref{plot1}) \cite{linde} 
appeared during the course of our work on this paper.

\subsection{The Spectrum for the Velocity Dependent Inflaton
Potentials from String Theory} 

The second class of modifications generically derived from string
theoretical inflationary models involves a velocity dependence in the
potential $V(\phi)\propto \frac{\dot\phi ^p}{\phi^q} + constant$ for
the inflaton field $\phi$. In string theory this term shows up and is
non negligible when branes move with respect to each other.

These models were treated in ~\cite{shiu-tye} by absorbing the velocity
dependent modification $Z(\phi)$ into the definition of the potential
and the slow roll parameters. The effective action for the inflaton in
this case is  
\be
\Gamma (\phi) = \int d^4 x \left (-a^3 V(\phi) + \frac{1}{2}a^3
Z(\phi)\dot\phi^2 - \frac{1}{2}a Z(\phi)(\nabla\phi)^2 \right) \,.
\ee 
The reader can find all the details of the calculation of the
modified primordial spectrum in ~\cite{shiu-tye}. Here we will just
report their expression for $\delta_H$ in terms of the unmodified
spectrum $\delta^0_H$, (the latter is obtained by the limit
$Z(\phi)\rightarrow 1$), and the stringy modification term $Z(\phi)$,
\be 
\delta_H = \delta^0_{H} Z(\phi) \,.  
\ee 
Therefore the running of
the spectral index is given by 
\be \delta n_S = n_S - n^0_{S} = 2
\frac{d \ln Z}{d \ln k} \,.  
\ee Clearly by comparing this class of
velocity dependent modifications with the modified Friedmann equation
of the previous subsection, {\it the effect of $Z(\phi)$ in the
spectrum is identical to that of $g(\phi)$}.  The running of the
spectral index is such that if the slow roll conditions persist in the presence of stringy modifications, $|\delta n_S| < 1$. Notice that $\delta n_S$ is a function of
the energy scale $k$ and runs with $k$ through its dependence on
$V$,$V^\prime$, $Z$, $Z^\prime$, $f$, $f^\prime$. For example, if we
are deep into the regime of new 
physics $V > E_c$, it is possible that these modifications are strong
enough to break the slow-roll conditions thus one would get a
different $|\delta n_S| \geq 1$ answer, if backreaction of stringy terms
violates the slow-roll regime as we show below with two examples. 

Since we are interested in the k-dependence of the modifications in
the primordial spectrum one can do the same algebraic tricks in order to express the
modifications in the spectrum as \be Z(\phi)^2 =
\left(\frac{k}{k_S}\right)^{\delta n_S} \propto  [1 - e^{-
(\frac{k}{k_S})^{\delta n_S}} ] \,.  
\ee 
The authors of ~\cite{shiu-tye} made the important observation that one
possible signature of string models is the fact that tensor
perturbations are not affected by these modifications, therefore the
ratio of tensor to scalar perturbations will be different from the
unmodified inflationary models. This is due to the different origins
of tensor fluctuations arising from bulk modes versus scalar
perturbations arising from brane quantum fluctuation modes. In this
work we are offering a new way for detecting signatures of the string
theoretical brane world models, through the suppression effect they
induce on the power spectrum at very low ``l''.

\subsection{Examples}

{\bf Modifications of Friedmann equation for the Randall Sundrum type
of brane-worlds}
  
In this scenario~\cite{rs} the modification to the Friedmann equation
for the brane bound
observer~\cite{binetruy,flanagan,csaki,laura,chung} is \be H^2 =
{\Lambda_4 \over 3} + \left({8\pi \over 3 M_{P}^2} \right) \rho +
\left({4\pi \over 3 M_5^3}\right)^2 \rho^2 + {\epsilon \over a^4}, \ee
where $\epsilon$ is an integration constant that appears as a form of
``dark radiation'' that is not important during inflation.  We will
also assume that the bulk cosmological constant $\Lambda \sim -4 \pi
\sigma^2/3 M_5^3$, so that the three-dimensional cosmological constant
$\Lambda_4$ is negligible during the early universe\footnote{This
  fine-tuning is the usual cosmological constant problem, which is not
  addressed in this paper.}.  Hence the relevant correction to the
Friedmann equation during inflation is the quadratic term, \be
f[\rho^{\alpha}]= \rho^2/2\sigma \,, \ee so that 
\be 
\label{eq:rho2}
H^2 = {8 \pi
  \over 3 M_{P}^2} \rho [1 + \rho/2\sigma]\,, 
\ee 
with the cutoff
scale given by the brane tension, 
\be 
E_c \sim |\sigma| .
\ee
 Notice that
in the absence of bulk fields, $\sigma < 0$ in order to ensure
stability. 
As can be seen from
Eqn.~(\ref{eq:denom}) we have a suppression of power when the
modification term is significant at early times. One recovers the
conventional spectrum at later times when the modification term
becomes negligible.

\be \left\{ \begin{array}{ll} \delta_H < \delta_H^{(0)} &
\mbox{when}\;\;\;\; \rho^2/|\sigma| \ge \rho \,,\\ & \\ \delta_H
\rightarrow \delta_H^{(0)} & \mbox{when} \;\;\;\;\rho^2/|\sigma| < \rho \,.
\end{array}
\right.  
\ee 

Under the slow roll conditions $\rho\propto V$,
Eqn.(\ref{eq:rho2}) becomes
\be 
\label{eq:slowroll}
H^2 = {8 \pi
  \over 3 M_{P}^2} V \, \left[1 - {V \over 2|\sigma|}\right]\,, 
\ee 
and the expression for $\delta_{H}^2$ can be derived by using
Eqns.~(\ref{gphi} -\ref{eq:correction}):  
\be 
\delta n_{S}(k) =
\frac{ -V^{\prime 2}}{ 3 H^2 (V - 2 |\sigma|  )} 
= \frac{9\dot \phi^2}{(2 |\sigma| -V )}
\,.  
\ee
We note that an equivalent expression to Eqn.~(\ref{eq:denom}) has
been derived previously ~\cite{maartens} in the slow roll regime.
\be
\delta_H^2 \sim \bigl({512 \pi \over 75 M_{P}^6}\bigr)
{V^3 \over {V^\prime}^2} \left[1 - {V \over 2 |\sigma|}\right]^3 .
\label{bruce}
\ee

However, the slow roll assumption breaks down at early times due
to the new physics modifications. In this example, if we extrapolate
backwards to a value $\phi_*$ of the inflaton field at which
$V(\phi_*) = 2|\sigma|$, we have $\delta n_S \rightarrow \infty$ and
$\delta_{H}^{2} \rightarrow 0$. Therefore near the critical point
$\phi_*$, the calculation for the spectrum should be done by using
Eqn.~(\ref{eq:denom}) instead of Eqn.~(\ref{bruce}) since the
approximation $\dot \phi \simeq -V^\prime / 3H$ breaks down.

At the high energy scales (early times) at which the slow roll
approximation breaks down, the energy density is dominated by the
kinetic energy of the inflaton, and density fluctuations are not
produced.  So this effect is imprinted in low $l$ since these modes
are the first to cross the horizon at the initial time. However, we
need to translate the suppression of these wavenumbers into the
``spreading'' that this feature has over the range of ``l's'' in the
present day power spectrum, $P(k)$, as seen in Fig.2.

The magnitude of suppression is of course model dependent and it is a
function of our assumptions about the modification term $V^2
/(2\sigma)$ and the energy scales for $V$, $|\sigma|$ which we took to
be $GUT$ scale here.
 
For the sake of concreteness, let us assume the following inflaton
potential ~\cite{maeda,stewart}, 
\be V = V_0 e^{-\lambda\phi} \,,
\ee 
(where $\phi = \Phi / \sqrt{8\pi} M_{P}$ is the inflaton field given
in Planck units in order to be dimensionless).  We shift
the field so that its initial value at 60 efolds before the end
of inflation is at $\phi_i=0$.
Then the solution for
the inflaton field in the presence of modifications 
becomes 
\be
\label{eq:inflaton}
\phi =
\frac{\lambda}{(1 - \frac{V_0}{2|\sigma|})} 
\ln(\frac{a H} {a_i H_i})_{k
  = aH} \sim - \lambda \,\, 
\Delta N \,\,{1 \over (1 - {V_0 \over 2 |\sigma|})} \,,
\ee
where 
\be
\Delta N = N_I(k) - N_I(k_s) = \ln({k_s \over k}) \,,
\ee
and $N_I(k)$ is the number of efoldings before the end of inflation
at which perturbations on scale $k$ leave the horizon.
The spectral index obtained in the usual exponential inflation
case without modified Friedmann equations is
given by $n^0_S - 1 = -\lambda^2$. 

By replacing the field solution we obtain 
\bea 
\label{eq:modg}
\tilde{g}(\phi) &=& [1 + \frac{V}{2\sigma} ]^3 \, = \, [1 -
\frac{V_0}{2|\sigma|} \, (\frac{k}{k_S})^{\frac{(n^0_S -1)}{(1-
    \frac{V_0}{2|\sigma|})} } ]^3\,,
\label{eqrs}  
\eea 
and
\be
\label{eq:moddelta}
\delta_H^2 \sim 8 \pi \bigl({512 \pi \over 75 M_{P}^4}\bigr) \, \, 
{V_0 \over {\lambda}^2} \left[ \frac{k}{k_S}  \right]^{\frac{(n^0_S -1)}{(1-
\frac{V_0}{2|\sigma|})} } \, \, {\tilde{g}(\phi)} .
\label{katie}
\ee
which produces the suppression feature in the low ``l'' regime,
corresponding to $V_0 \simeq |\sigma|$, i.e the regime where new
physics becomes important. 
Here, $k_S$ corresponds to the scale of perturbations
produced 60 efoldings before the end of inflation, 
which corresponds to the horizon size today,
$k_S \sim (4000 Mpc)^{-1}$.

In Figure 2, we present the results obtained using the CMBFAST code
with the modifications of Eqns.~(\ref{eq:modg}) and
(\ref{eq:moddelta}) for two different values of $\lambda$ and of
$k_S$, with the height of the potential determined by $|\sigma| \sim
M_{GUT} \sim 10^{16}$GeV.  In the plot, we have chosen to normalize
all the curves to the value of the WMAP $l=17$ multipole for
comparison. In order to match the amplitude of the data, we choose
values of $\lambda^2 \sim 0.1$ as shown in the plot.  We see that the
Doppler and higher peaks can be essentially unaffected by these
modifications, while the low $l$ modes are substantially suppressed.
Hence the WMAP data is better fit by these modifications than by a
standard $\Lambda$-CDM model.  We have not performed a quantitative
study in this paper of how good the fit is to the data, which for a
particular model may be the subject of future work; it is our goal
here to display the possible role of new physics in the suppression of
low ``l'' modes.

In order to illustrate the role of modifications, let us start in the
slow roll regime, where Eqn.~(\ref{eq:slowroll}) is valid, and go
backwards in time (up the potential).  At first, as we climb up the
potential, the value of the Hubble constant $H$ increases.  Then, when
we reach the value of $\phi$ at which $V(\phi) = |\sigma|$, we see that
we have reached a maximum for $H$, i.e., $\delta H/ \delta V = 0$.  In
other words, $H$ is maximized for $V = |\sigma|$.  When we continue to
climb up the hill, $H$ decreases towards zero, and the slow roll
approximation begins to fail as the kinetic energy of the inflaton
becomes more and more important.  At $\rho(\phi_*) = 2|\sigma|$ the
spectrum $\delta_H \rightarrow 0$, $n_S \rightarrow \infty$, thus
signalling a completely stringy regime and the slow roll conditions
being badly violated.  Notice that modifying the Hubble parameter, $H$
through the ``stringy terms'', changes the kinetic energy of the
inflaton, since $H$ play the role of a friction term for the field.

Let us parametrize  the kinetic energy of the inflaton as $K
=\frac{\dot \phi^2}{2} = \alpha V$.  Slow roll requires
that $\alpha \leq 1/2$. Then the inflaton
energy density is $\rho(\phi) = (1 + \alpha) V(\phi)$. The
value of the field $\phi_*$ for which the modified Hubble parameter
vanishes $H=0$, is given by the solution $\phi_* \simeq
\frac{1}{\lambda} \ln [(\frac{2|\sigma|}{V_0})(\frac{1}{1+\alpha_*})],
(1+\alpha_*) \leq 3/2$. At $\phi =\phi_*$ where $H=0$ the inflaton has a
lot of kinetic energy. But when the field rolls at some $\phi_i$ given
by $\rho(\phi_i) \simeq O(|\sigma|)$ we have the start of the
slow-roll regime with the kinetic energy of the inflaton decreasing to
less than $V(\phi_i)/2$. For field values between $\phi_i > \phi
>\phi_*$  the ``friction term'' $H$  starts increasing and the kinetic
energy of the inflaton decreases. The point where the field acquires
the value $\phi_i \simeq \frac{1}{\lambda} \ln
[(\frac{|\sigma|}{V_0})(\frac{1}{1+\alpha_i})] $ is a maximum 
for $H^2 (\phi)$. Clearly $\alpha_i \ll \alpha_* \leq 1/2$. In the
region of the validity of new  
physics, due to the modified friction term $H$ for the field, i.e. for
$\phi_i \geq \phi \geq \phi_*$, the inflaton was in a ``fast-roll''
since its kinetic energy was large due to the highly suppressed
modified ``friction term'' $H$. Therefore all long wavelength
perturbations are highly suppressed in this regime. However as the
modified Hubble parameter increases towards $\phi = \phi_i$ and
attains its maximum at $\phi_i$ then the kinetic energy of the
inflaton starts decreasing until it reaches a minimum at $\phi_i$. At
$\phi_i$ its potential energy $V(\phi_i) \simeq |\sigma|$ dominates over
the kinetic term and the inflaton enters the slow-roll regime of a
conventional inflationary period, even in the presence of a modified
Hubble parameter. The Hubble parameter and $V(\phi)$ slowly decrease
for the standard model regime of the slow roll, $\phi_{end} \geq \phi
\geq \phi_i$ thus rendering the stringy correction terms insignificant
in this range of $\phi$. 

In short, modified Friedmann equations of RSI provide a possible
explanation of suppressed large scale power in the WMAP data,
for parameter choices such as those shown in Figure 2.

\begin{figure} 
\epsfysize=9cm \epsfxsize=9cm \hfil \epsfbox{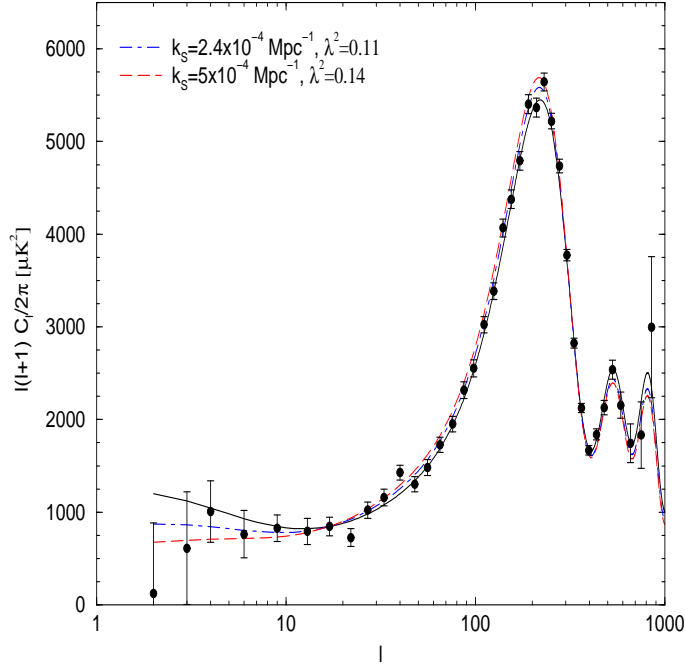} \hfil
\caption{\footnotesize CBR spectrum of anisotropies with the modified
primordial spectrum for the Randall Sundrum example,
Eqn.~(\ref{eqrs}), with $V_0 \simeq |\sigma| \sim m_{GUT}$; $k_S=0.0005\,
Mpc^{-1}$ (dashed line) and $k_S=0.00024 \, Mpc^{-1}$ (dot-dashed
line). We also show for comparison the curve for the standard     
spectrum Eq (\ref{unmodprimordial}), with $n_S^{(0)}=0.99$ (solid line). 
Cosmological parameters as given by the best fit model
of the WMAP collaboration.}
\label{plot2}
\end{figure}

{\bf Running Spectral Index}

The WMAP data show some evidence that the spectral index $n_s$ runs
(changes as a function of the scale $k$ at which it is measured) from
$n_s>1$ (blue) on large scales to $n_s<1$ (red) on small scales.
Chung, Shiu and Trodden \cite{cst} proposed that such a running index
may be obtained in inflationary models if the slope of the potential
reaches a minimum in the regime where the spectral behavior changes;
such a potential must be carefully chose.  In another proposal
\cite{kyy,feng2}, an inflationary model motivated by supergravity was
proposed as an explanation of the running spectral index, in which a
period of hybrid inflation is followed by a second period of
inflation.  We here propose an alternate possibility for explaining
the running spectral index.

We will see that the density perturbations from modified Friedmann
equations in
Eqn.(\ref{eq:moddelta}) (although not described in terms
of an overall spectral index) give rise to a spectrum that is 
blue on large scales and red on small scales, in agreement
with the WMAP data.  We find that $\delta_H^2$ is an increasing
function of $k$ for small scales, and a decreasing function for
large scales.  We find that $\delta_H^2$ reaches a maximum
at 
\be
k_{max}/k_s = \left[{|\sigma| \over 2 V_0} \right]^{{1 \over \tilde{n}
-1} }
\ee
where
\be
\tilde{n}-1 = {\frac{(n^0_S -1)}{(1-
    \frac{V_0}{2|\sigma|})}} .
\ee
For $V_0 = |\sigma|$ and taking $\lambda^2 = 0.11$ as in Fig.(2),
we find that $k_{max}/k_s = 23$ so that the maximum power is
on scales of roughly 180 Mpc.  The power decreases as one moves
away from the maximum in either direction.  Hence, compared
to a flat spectrum, the spectral index is shifted to the red on
small scales and blue on large scales.

Hence modified Friedmann equations have the capability of
explaining not only the suppression of power on the scales
of the quadrupole, but also the running of the spectral index
observed in the data.

{\bf Velocity dependent potentials from string theoretical
inflationary models}

As mentioned in Section 1, the cosmological consistency conditions for
this class of modifications were treated in great detail in
~\cite{shiu-tye}. In our work we are interested to search for
possible low ``l''
 signatures that this class of models may give rise
to, thus presenting this class of stringy modifications as a second example. In order to illustrate our point, let us consider the following
velocity dependent modification as a representative example for this
class 
\be Z(\phi) = 1 - V(\phi)/E_c \,.  \ee
Post Newtonian corrections of higher dimensional gravity are also
expected to give rise to such modification terms.  In this case the
correction power is given by 
\be \delta n_S = (2/Z) (d Z / d
\phi)(\dot \phi /H) = \frac{2 V^{\prime 2}}{3 H^2 E_c (1 - \frac{V}{E_c} )^2}
\,.  \ee 
In the regime 
where the slow roll condition is valid, $\delta n_S < 1$ and therefore we can
approximate the modification term in the primordial spectrum by an
exponential. The modified primordial spectrum will then be given by
Eqn.~(\ref{eq:exp}) with $\dot \phi \simeq \frac{- V^\prime}{3 H Z}$. 

If we now take $V = V_0 e^{\lambda\phi}$ as in the previous case, this
class of modifications is another example whereby we again obtain a
suppression of the signal at low ``l'' through

\bea
\delta n_S &=& 2 \, \lambda^2 \, 
(V_0/E_c) (\frac{k}{k_S})^{\frac{\lambda^2}{[1 - (V_0/E_c]}} \,\,\frac{1}{[1 -
(V_0/E_c) (\frac{k}{k_S})^{\frac{\lambda^2}{[1 - (V_0/E_c]}} ]^2} \,,
\eea

Again this correction term blows up at $V_0 = E_C$ thus violating
slow-roll and introducing a cutoff on the spectrum $\delta_H \simeq 0$
at the energy scale $V(\phi_*) = E_c$. Hence the period of fast roll
is due to the modified friction term $H$ for the inflaton at high
energies, $V \simeq O(E_c)$, and it comes to an end when the Hubble
parameter, due to the rolling field, goes through its saddle point
extremum at $V(\phi_i) \simeq \frac{E_c}{2}, \phi_i \simeq
\frac{1}{\lambda} \ln (E_c /2 V_0 )$. This is also the point where the
kinetic energy of the inflaton drops by more than half its potential
energy and the inflaton enters into the slow roll phase from this
point onwards. Also since $V(\phi_{end}) \ll V(\phi_i) \ll V(\phi_*)$
the correction terms soon become small when the field enters the phase
of the standard physics regime, for $\phi \leq \phi_i$.

It should be noticed that the initial condition and the start of the
slow-roll regime at $\phi_i$, in both examples, result from and are
fixed by the stringy correction term of the new physics because this
modification determines the saddle point location for the ``friction
term'' of the field, namely the Hubble parameter $H$. Thus the
model-dependence can not be avoided in the absence of knowledge of the
fundamental theory of new physics. However, observational data ought
to discriminate between the various theoretical models for the high
energy regime.

\section{Summary}
The new results for the suppression of the low order multipoles in the
microwave background spectrum, measured by the Wilkinson Microwave
Anisotropy Probe mission, are very intriguing.  They may give us clues
towards new physics that determines the initial conditions for
inflation.  As examples of new physics that may be responsible for
this deviation we investigate modified Friedmann equations and
velocity dependent potentials, that may have resulted from a deeper
underlying theory, for example string theory.  Using the CMBFAST code
we see that indeed fundamental new physics can give rise to the
feature observed by WMAP at low ``l''. However the type of feature
produced is determined by the model chosen for the stringy
modifications at high energies, as we illustrated with the examples of
Sec. 2 above. The important issue, as we have shown in this work, is
that the low ``l'' signatures in the CBR spectrum would offer another
piece of evidence for testing string cosmology and our assumptions
about the initial conditions of inflation originating from new
physics.  In addition, these modifications to the basic physics have
the capability of explaining not only the suppression of power on the
scales of the quadrupole, but also the running of the spectral index
observed in the data.  It is exciting that ideas
about new physics and the initial conditions may be testable and
within observational reach.

In some cases, the stringy modifications may become important once
more, at very recent times. An example is the class of string inspired
models that contain late time effects in the form of dark
energy~\cite{tp,cardassian, kogan, gregory, dvali}. A very interesting
relation exists in that case between the two apparent coincidences
around redshifts $z \leq 1$, dark energy domination and the low ``l''
suppression; however, we will report on this scenario in a separate
publication.

\bigskip
\bigskip
\bigskip
{\it \bf ACKNOWLEDGMENTS}: We would like to thank S.Carroll, C. 
Csaki, P. Greene, W. Kinney, J. Liu, M. Trodden, G. Shiu, G. Starkman and
R. Sorkin, for helpful conversations.  LM's work is supported is
supported in part by the U.S. Department of Energy under grant
DE-FG02-85ER40231 and by the National Science Foundation under grant
PHY-0094122.   
KF acknowledges support from the Department of Energy through a grant
at the University of Michigan; KF also thanks the Michigan Center for
Theoretical Physics for support.
\\

\end{document}